\documentclass[11pt,a4paper]{article}

\usepackage{amssymb}
\usepackage[dvips]{graphicx}
\usepackage{bm}

\begin{document}

\title{One-loop polarization operator of the quantum gauge superfield for ${\cal N}=1$ SYM regularized by higher derivatives}

\author{A.E.Kazantsev, M.B.Skoptsov, and K.V.Stepanyantz\\
{\small{\em Moscow State University}}, {\small{\em  Physical
Faculty, Department  of Theoretical Physics}}\\
{\small{\em 119991, Moscow, Russia}}}

\maketitle

\begin{abstract}
We consider the general ${\cal N}=1$ supersymmetric gauge theory with matter, regularized by higher covariant derivatives without breaking the BRST invariance, in the massless limit. In the $\xi$-gauge we obtain the (unrenormalized) expression for the two-point Green function of the quantum gauge superfield in the one-loop approximation as a sum of integrals over the loop momentum. The result is presented as a sum of three parts: the first one corresponds to the pure supersymmetric Yang--Mills theory in the Feynman gauge, the second one contains all gauge dependent terms, and the third one is the contribution of diagrams with a matter loop. For the Feynman gauge and a special choice of the higher derivative regulator in the gauge fixing term we analytically calculate these integrals in the limit $k\to 0$. In particular, in addition to the leading logarithmically divergent terms, which are determined by integrals of double total derivatives, we also find the finite constants.
\end{abstract}

\section{Introduction}
\hspace*{\parindent}

\begin{figure}[h]
\begin{picture}(0,2.1)
\put(6.8,0){\includegraphics[scale=0.18]{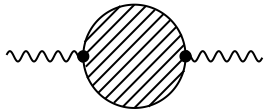}}
\end{picture}
\caption{The diagrams contributing to the two-point Green function of the quantum gauge superfield.}\label{Figure_Effective_Diagram}
\end{figure}

Investigation of higher loop quantum corrections in ${\cal N}=1$ supersymmetric gauge theories is an interesting and complicated problem needed for doing precise calculations in quantum field theory models \cite{Mihaila:2013wma}. Moreover, explicitly calculating quantum corrections one can verify some general statements about their structure. For example, the explicit three- \cite{Jack:1996vg,Jack:1996cn} and four-loop \cite{Harlander:2006xq} calculations  demonstrated that the NSVZ $\beta$-function \cite{Novikov:1983uc,Jones:1983ip,Novikov:1985rd,Shifman:1986zi} is not valid if the calculations are made in the $\overline{\mbox{DR}}$-scheme, due to the scheme-dependence of the NSVZ relation \cite{Kutasov:2004xu,Kataev:2014gxa}. Therefore, to obtain the NSVZ scheme, in each order it is necessary to make a special finite renormalization which relates it to the $\overline{\mbox{DR}}$ scheme \cite{Jack:1996vg,Jack:1996cn,Jack:1998uj}. The multiloop calculations made with the higher derivative regularization \cite{Slavnov:1971aw,Slavnov:1972sq} in the supersymmetric version \cite{Krivoshchekov:1978xg,West:1985jx} allowed to understand how to construct the general all-order prescription for obtaining the NSVZ scheme in supersymmetric theories both in the Abelian \cite{Kataev:2013eta,Kataev:2013csa,Kataev:2014gxa} and non-Abelian \cite{Stepanyantz:2016gtk} cases. These calculations (see, e.g., \cite{Pimenov:2009hv,Stepanyantz:2011bz,Kazantsev:2014yna,Buchbinder:2014wra,Buchbinder:2015eva,Aleshin:2016yvj,Shakhmanov:2017soc}) demonstrate that the $\beta$-function in supersymmetric theories regularized by higher covariant derivatives is given by integrals of total derivatives \cite{Soloshenko:2003nc} and even double total derivatives \cite{Smilga:2004zr}.\footnote{Such a factorization into integrals of total derivatives does not take place in the case of using the dimensional reduction \cite{Siegel:1979wq}, see \cite{Aleshin:2015qqc,Aleshin:2016rrr}.} At present, this has also been proved in all loops in the Abelian case \cite{Stepanyantz:2011jy,Stepanyantz:2014ima}. The similar factorization into double total derivatives has been proved for the Adler $D$-function \cite{Adler:1974gd} in ${\cal N}=1$ SQCD \cite{Shifman:2014cya,Shifman:2015doa} and for the photino mass renormalization in the softly broken ${\cal N}=1$ SQED \cite{Nartsev:2016nym}. Such a structure of loop integrals leads to the NSVZ (or NSVZ-like \cite{Hisano:1997ua,Jack:1997pa,Avdeev:1997vx}) relations for the renormalization group (RG) functions defined in terms of the bare coupling constant. Consequently, the NSVZ (or NSVZ-like) scheme for the RG functions defined in terms of the renormalized coupling constant \cite{Kataev:2014gxa,Kataev:2013eta,Kataev:2013csa,Stepanyantz:2016gtk,Nartsev:2016mvn} appears to be the higher covariant derivative regularization supplemented by the subtraction scheme in which only powers of $\ln\Lambda/\mu$ are included into the renormalization constants, where $\Lambda$ is the dimensional parameter of the regularized theory and $\mu$ is a normalization point. In some sense, this scheme is analogous to the minimal subtractions, so that we will call it HDMS. However, to explicitly verify that this scheme really coincides with the NSVZ scheme, one has to calculate the $\beta$-function at least in the three-loop approximation, where the scheme dependence becomes essential. This problem is very complicated from the technical point of view due to the large number of supergraphs which should be considered. Nevertheless, sometimes it is possible to simplify some parts of the multiloop calculations. Say, there is a large number of multiloop graphs which contain subdiagrams of the structure presented in Fig. \ref{Figure_Effective_Diagram}.

In this paper we calculate the sum of the one-loop superdiagrams which have such a structure for the general ${\cal N}=1$ supersymmetric Yang--Mills theory with matter, regularized by the BRST-invariant version of the higher covariant derivative regularization. In other words, we calculate the two-point Green function of the quantum gauge superfield in the one-loop approximation. For this purpose, in Sect. \ref{Section_Regularization} we introduce the theory and describe the regularization procedure. In Sect. \ref{Section_Result} we present the diagrams contributing to the considered Green function and the result of the calculation given by Eq. (\ref{General_Result}). This equation is the main result of this paper. It consists of three parts: the expression for the pure supersymmetric Yang--Mills theory in the minimal gauge, the gauge dependent part, and the matter contribution. For the Feynman gauge $\xi_0=1$ and a special higher derivative regulator in the gauge fixing term the integrals in Eq. (\ref{General_Result}) are calculated in the limit $k/\Lambda\to 0$. The details of this calculation are described in the Appendix, where we find both logarithmically divergent terms and finite constants.

\section{${\cal N}=1$ SYM and the higher derivative regularization}
\hspace*{\parindent}\label{Section_Regularization}

In this paper we consider the massless ${\cal N}=1$ supersymmetric gauge theory with a simple gauge group $G$ and matter superfields in a certain representation $R$. The most convenient way to describe this theory is to use the ${\cal N}=1$ superfields, see, e.g., \cite{West:1990tg,Buchbinder:1998qv}. Then the action can be written in the form

\begin{equation}
S = \frac{1}{2e_0^2} \mbox{tr}\,\mbox{Re} \int d^4x\, d^2\theta\, W^a W_a + \frac{1}{4} \int d^4x\,d^4\theta\, \phi^{*i} \big(e^{2V}\big)_i{}^j \phi_j
+ \Big(\frac{1}{6} \int d^4x\, d^2\theta\, \lambda_0^{ijk} \phi_i \phi_j \phi_k +\mbox{c.c.}\Big),
\end{equation}

\noindent
where $V(x,\theta)$ is the gauge superfield and $\phi_i$ denotes chiral matter superfields in the representation $R$ of the gauge group. In our notation, the supersymmetric gauge superfield strength is $W_a \equiv \bar D^2\big(e^{-2V} D_a e^{2V}\big)/8 = e_0 W^A t^A$, and the generators of the fundamental representation are normalized by the condition $\mbox{tr}(t^A t^B) = \delta^{AB}/2$. (Certainly, in the action for the matter superfields $V = e_0 V^A T^A$, where $T^A$ are the generators of the representation $R$.) The subscript $0$ indicates bare couplings.

In this paper we will use the same version of the higher covariant derivative regularization as in Ref. \cite{Aleshin:2016yvj}. This regularization is convenient, because it does not break the BRST-invariance of the gauge-fixed theory \cite{Becchi:1974md,Tyutin:1975qk}.\footnote{Certainly, it is possible to use simpler versions of the higher derivative regularization breaking the BRST invariance, see, e.g., \cite{Pimenov:2009hv,Stepanyantz:2011bz}. However, it is highly desirable that a regularization does not break as many symmetries as possible in order that the corresponding Slavnov--Taylor identities take place for the regularized theory.} Also it is convenient to use the background field method, which is introduced by making the quantum-background splitting

\begin{equation}
e^{2V} \to e^{\bm{\Omega}^+} e^{2V} e^{\bm\Omega},
\end{equation}

\noindent
where $\bm{\Omega}$ is related to the background gauge superfield $\bm{V}$ by the equation $e^{2\bm{V}} = e^{\bm{\Omega}^+} e^{\bm{\Omega}}$. Similarly, we introduce the gauge extension $\Omega$ of the superfield $V$ by the equation $e^{2V} = e^{\Omega^+} e^\Omega$. Then, following Ref. \cite{Aleshin:2016yvj}, we add the higher derivative term $S_\Lambda$ to the classical action, so that

\begin{eqnarray}
&& S + S_\Lambda =  \frac{1}{2e_0^2}\,\mbox{Re}\,\mbox{tr}\int d^4x\,
d^2\theta\, W^a \Big[e^{-\bm{\Omega}}
e^{-\Omega} R\Big(-\frac{\bar\nabla^2 \nabla^2}{16\Lambda^2}\Big) e^{\Omega} e^{\bm{\Omega}}\Big]_{Adj} W_a + \frac{1}{4} \int d^4x\, d^4\theta\,\phi^+ \qquad\nonumber\\
&& \times e^{\bm{\Omega}^+} e^{\Omega^+} F\Big(-\frac{\bar\nabla^2
\nabla^2}{16\Lambda^2}\Big) e^{\Omega} e^{\bm{\Omega}} \phi + \Big(\frac{1}{6} \int d^4x\, d^2\theta\, \lambda_0^{ijk}\phi_i \phi_j \phi_k +\mbox{c.c.}\Big),
\end{eqnarray}

\noindent
where the functions $R$ and $F$ rapidly growing at infinity satisfy the conditions $R(0) = F(0)=1$ and

\begin{equation}
W_a = \frac{1}{8} \bar D^2\Big(e^{-\bm{\Omega}} e^{-2V} e^{-\bm{\Omega}^+} D_a\big(e^{\bm{\Omega^+}} e^{2V} e^{\bm{\Omega}}\big)\Big).
\end{equation}

\noindent
The covariant derivatives are defined as

\begin{equation}
\nabla_a = e^{-\Omega^+} e^{-\bm{\Omega}^+} D_a e^{\bm{\Omega}^+}
e^{\Omega^+}; \qquad \bar\nabla_{\dot a} = e^{\Omega}
e^{\bm{\Omega}} \bar D_{\dot a} e^{-\bm{\Omega}} e^{-\Omega}.
\end{equation}

\noindent
The gauge fixing term invariant under the background gauge transformations is chosen in the form

\begin{equation}
S_{\mbox{\scriptsize gf}} = -\frac{1}{16 \xi_0 e_0^2}\mbox{tr} \int
d^4x\,d^4\theta\,\bm{\nabla}^2 V K\Big(-\frac{\bm{\bar\nabla}^2
\bm{\nabla}^2}{16\Lambda^2}\Big)_{Adj}\bm{\bar \nabla}^2 V,
\end{equation}

\noindent
where

\begin{equation}
\bm{\nabla}_a = e^{-\bm{\Omega}^+} D_a e^{\bm{\Omega}^+}; \qquad \bm{\bar\nabla}_{\dot a} = e^{\bm{\Omega}} \bar D_{\dot a} e^{-\bm{\Omega}},
\end{equation}

\noindent
so that the corresponding Faddeev--Popov action is

\begin{eqnarray}
&& S_{\mbox{\scriptsize FP}} = \frac{1}{e_0^2} \mbox{tr} \int
d^4x\,d^4\theta\, \left(e^{\bm{\Omega}}\bar c e^{-\bm{\Omega}} +
e^{-\bm{\Omega}^+}\bar c^+ e^{\bm{\Omega}^+}\right)\nonumber\\
&&\qquad\qquad\qquad\quad \times \Big\{
\Big(\frac{V}{1-e^{2V}}\Big)_{Adj} \Big(e^{-\bm{\Omega}^+} c^+
e^{\bm{\Omega}^+}\Big)  + \Big(\frac{V}{1-e^{-2V}}\Big)_{Adj}
\Big(e^{\bm{\Omega}} c e^{-\bm{\Omega}}\Big)\Big\}.\qquad
\end{eqnarray}

\noindent
Also it is necessary to introduce the Nielsen--Kallosh ghosts. They interact only with the background gauge superfield and are not essential in this paper. Their action $S_{\mbox{\scriptsize NK}}$ can be found in \cite{Aleshin:2016yvj}.

For regularizing one-loop divergences which remain after adding $S_\Lambda$ we insert into the generating functional the Pauli--Villars determinants \cite{Slavnov:1977zf,Faddeev:1980be}. According to \cite{Aleshin:2016yvj}, the one-loop divergences introduced by the gauge and ghost superfields are cancelled if one introduces three chiral (commuting) Pauli--Villars superfields $\varphi_\alpha$ with $\alpha=1,2,3$ in the adjoint representation of the gauge group,

\begin{eqnarray}\label{Pauli--Villars_Actions}
&&\hspace*{-6mm} S_\varphi = \frac{1}{2e_0^2} \mbox{tr} \int
d^4x\,d^4\theta\, \Big(\varphi_1^+ \Big[e^{\bm{\Omega}^+}
e^{\Omega^+} R\Big(-\frac{\bar\nabla^2 \nabla^2}{16\Lambda^2}\Big)
e^{\Omega} e^{\bm{\Omega}}\Big]_{Adj}\varphi_1 + \varphi_2^+
\Big[e^{\bm{\Omega}^+} e^{2V}
e^{\bm{\Omega}}\Big]_{Adj}\varphi_2 \nonumber\\
&&\hspace*{-6mm} + \varphi_3^+ \Big[e^{\bm{\Omega}^+} e^{2V}
e^{\bm{\Omega}}\Big]_{Adj}\varphi_3\Big) +
\frac{1}{2e_0^2}\mbox{tr}\Big(\int d^4x\,d^2\theta\,M_\varphi
\left(\varphi_1^2 + \varphi_2^2 + \varphi_3^2\right)
+\mbox{c.c.}\Big).
\end{eqnarray}

\noindent
To cancel the one-loop divergences coming from the matter loop, we introduce the chiral (commuting) Pauli--Villars superfield $\Phi_i$ in a certain representation $R_{\mbox{\scriptsize PV}}$ of the gauge group with the action

\begin{equation}
S_\Phi = \frac{1}{4} \int
d^4x\,d^4\theta\,\Phi^{*i} \Big[e^{\bm{\Omega}^+} e^{\Omega^+}
F\Big(-\frac{\bar\nabla^2 \nabla^2}{16\Lambda^2}\Big) e^{\Omega}
e^{\bm{\Omega}}\Big]_i^{\ \ j} \Phi_j + \Big(\frac{1}{4} \int
d^4x\,d^2\theta\, M^{ij} \Phi_i \Phi_j +\mbox{c.c.}\Big),
\end{equation}

\noindent
where we assume the existence of the gauge invariant mass term such that $M^{ji} M^*_{kj} = M^2\, \delta_k^i$.

Then, we insert into the generating functional

\begin{equation}
\mbox{Det}(PV,M_\varphi)^{-1} \cdot \mbox{Det}(PV,M)^{c} = \int D\varphi_1 D\varphi_2 D\varphi_3 \exp\left(i S_\varphi \right) \cdot \Big(\int D \Phi \exp\left(i
S_\Phi\right)\Big)^{-c},
\end{equation}

\noindent
where $c = T(R)/T(R_{\mbox{\scriptsize PV}})$. It is important that the masses of the Pauli--Villars superfields should be proportional to the parameter $\Lambda$ which appears in the higher derivative term and the coefficients of the proportionality are independent of the coupling constant,

\begin{equation}
M_\varphi \equiv a_\varphi \Lambda;\qquad M \equiv a \Lambda.
\end{equation}

\section{One-loop two-point Green function of the quantum gauge superfield}
\hspace{\parindent}\label{Section_Result}

In this paper we find the two-point Green function of the quantum gauge superfield for the theory described in the previous section in the one-loop approximation. This may considerably simplify some multiloop calculations, because in calculating higher order corrections to various Green functions one often encounters graphs containing the subdiagrams of the structure presented in Fig. \ref{Figure_Effective_Diagram}. Certainly, it is necessary to distinguish the two-point Green functions of the background gauge superfield and of the quantum gauge superfield. The former Green function is evidently transversal due to the unbroken background gauge invariance. The quantum corrections to the two-point Green function of the quantum gauge superfield are also transversal, but in the tree approximation it is necessary to take into account that the gauge fixing action is not transversal.\footnote{The transversality of the quantum corrections in this case can be proved with the help of the Slavnov--Taylor identities \cite{Taylor:1971ff,Slavnov:1972fg} and the Schwinger--Dyson equations for ghosts.} In the supersymmetric case this implies that the part of the effective action corresponding to the two-point Green function of the quantum gauge superfield $V$ can be presented in the form

\begin{equation}\label{Dq_Definition}
\Gamma^{(2)}_V - S_{\mbox{\scriptsize gf}}^{(2)} = - \frac{1}{8\pi} \mbox{tr} \int \frac{d^4k}{(2\pi)^4}\, d^4\theta\, V(-k,\theta) \partial^2 \Pi_{1/2} V(k,\theta)\, d_q^{-1}\big(\alpha_0,\lambda_0,k^2/\Lambda^2\big),
\end{equation}

\noindent
where $\partial^2\Pi_{1/2} \equiv - D^a \bar D^2 D_a/8$ is the supersymmetric analog of the transversal projection operator. The coefficient in Eq. (\ref{Dq_Definition}) is chosen so that in the tree approximation (for the non-regularized theory) the function $d_q$ coincides with the bare coupling constant $\alpha_0$. Thus, quantum corrections to this function are encoded in the function $\Pi$ defined by the equation

\begin{equation}\label{Polarization_Operator}
d_q^{-1}(\alpha_0,\lambda_0,k^2/\Lambda^2) - \alpha_0^{-1} R(k^2/\Lambda^2) \equiv - \alpha_0^{-1} \Pi(\alpha_0,\lambda_0,k^2/\Lambda^2).
\end{equation}

\noindent
(The regulator $R$ is present in the action of the regularized theory and, therefore, should be included into this definition.) The function $\Pi$ can be called the polarization operator of the quantum gauge superfield in the supersymmetric case. To find this function in a certain approximation, it is necessary to calculate all corresponding 1PI diagrams of the structure presented in Fig. \ref{Figure_Effective_Diagram}.

If the polarization operator is known, then the effective propagator of the gauge superfield $V^A$ can be written as

\begin{equation}\label{Effective_Propagator}
2i\left(\frac{1}{\big(R -\Pi\big)\partial^2} - \frac{1}{16\partial^4}\Big(D^2 \bar D^2 + \bar D^2 D^2\Big)\Big(\frac{\xi_0}{K} - \frac{1}{R-\Pi}\Big)\right) \delta^8(x_1-x_2) \delta^{AB},
\end{equation}

\noindent
where $R = R(\partial^2/\Lambda^2)$, $K=K(\partial^2/\Lambda^2)$, and $\Pi = \Pi(\alpha_0,\lambda_0,\partial^2/\Lambda^2)$. With the help of this equation the calculations of multiloop diagrams containing the subdiagrams of the structure presented in Fig. \ref{Figure_Effective_Diagram} can be considerably simplified.

\begin{figure}[h]

\begin{picture}(0,4)

\put(2.5,2.4){\includegraphics[scale=0.4]{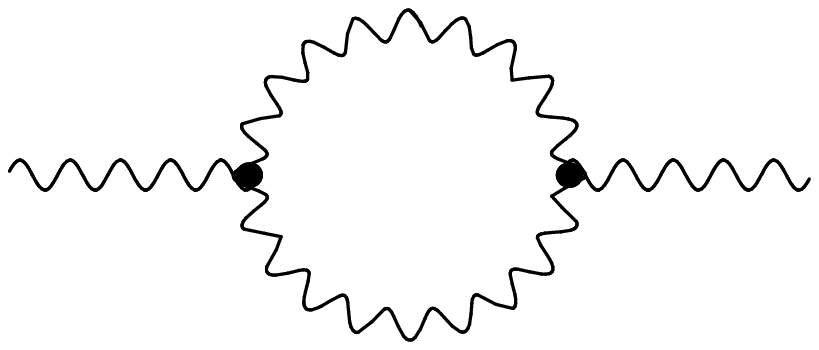}}

\put(2.9,0){\includegraphics[scale=0.4]{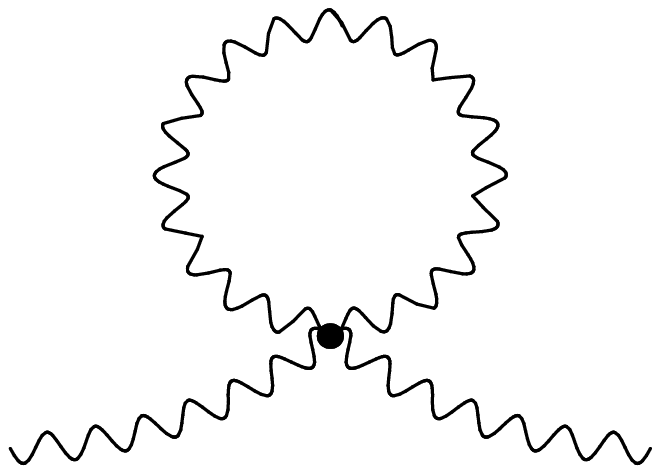}}

\put(6.4,2.43){\includegraphics[scale=0.4]{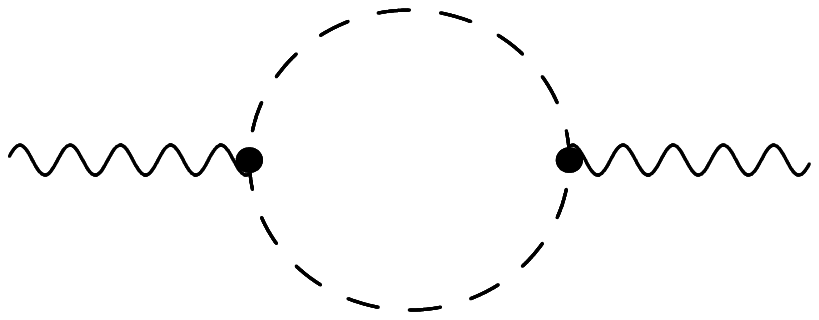}}

\put(6.7,-0.01){\includegraphics[scale=0.4]{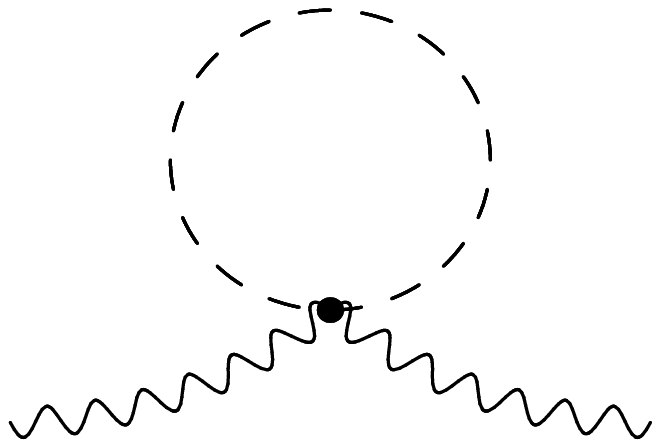}}

\put(10.3,2.4){\includegraphics[scale=0.4]{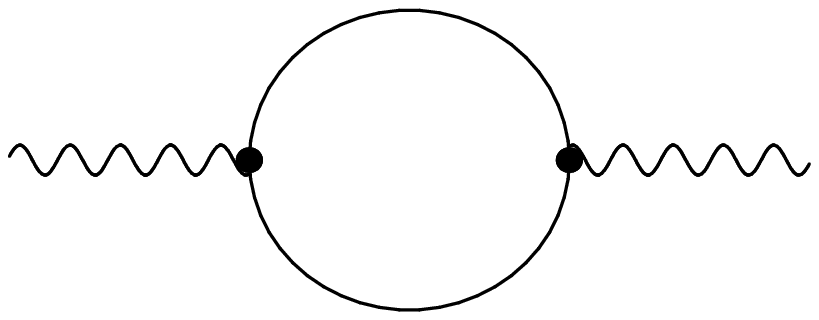}}

\put(10.7,-0.02){\includegraphics[scale=0.4]{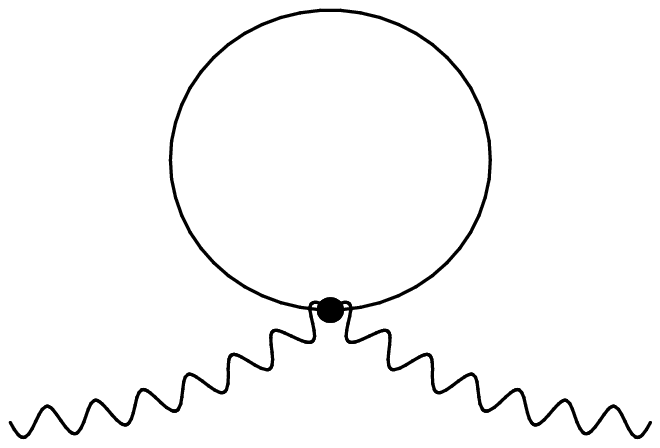}}

\end{picture}

\caption{One-loop contribution to the polarization operator of the quantum gauge superfield.}
\label{Figure_Polarization}

\end{figure}

In this paper we calculate the function (\ref{Polarization_Operator}) in the one-loop approximation. It is contributed to by the diagrams presented in Fig. \ref{Figure_Polarization}, where the wavy lines correspond to the quantum gauge superfield $V$, the dashed lines correspond to the Faddeev--Popov ghosts, and the solid lines correspond to the chiral matter superfields (including the Pauli--Villars ones).

After calculating the diagrams presented in Fig. \ref{Figure_Polarization} in the case of using the BRST-invariant version of the higher covariant derivative regularization described above we have obtained the following result for the (unrenormalized) polarization operator:

\begin{equation}\label{General_Result}
\Pi(\alpha_0,\lambda_0,k^2/\Lambda^2) = -8\pi\alpha_0 \Big(C_2\, f(k/\Lambda)+ C_2\, g(\xi_0,k/\Lambda) + T(R)\,h(k/\Lambda)\Big) + O(\alpha_0^2,\alpha_0\lambda_0^2),
\end{equation}

\noindent
where $k\equiv \sqrt{k^2}$ is the length of the Euclidean momentum $k_\alpha$. Here the contribution containing the function $f$ gives the result for the pure supersymmetric Yang--Mills theory in the minimal gauge ($\xi_0=1$ and $R(x) = K(x)$), the function $g$ is a sum of all gauge dependent terms, and the function $h$ includes all contributions of the chiral matter superfields $\phi_i$ and the corresponding Pauli--Villars superfields $\Phi_i$. These functions are given by the following Euclidean integrals:

\begin{eqnarray}\label{General_F}
&& f(k/\Lambda) = - \int\frac{d^4l}{(2\pi)^4} \Biggl[\frac{3}{2} \left(\frac{1}{l^2(l+k)^2}-\frac{1}{(l^2+M_{\varphi}^2)((l+k)^2+M_{\varphi}^2)}\right) -\frac{R_{l}-R_{k}}{R_{l}l^2}\left(\frac{1}{(l+k)^2}\right.\nonumber\\
&& \left.-\frac{1}{l^2-k^2}\right) -\frac{2}{R_l \big((l+k)^2-l^2\big)}\left(\frac{R_{l+k}-R_{l}}{(l+k)^2-l^2}-\frac{R'_{l}}{\Lambda^2}\right) + \frac{1}{R_{l}R_{l+k}} \left(\frac{R_{l+k}-R_{l}}{(l+k)^2-l^2}\right)^2  \nonumber\\
&& + \frac{2 R_{k}k^2}{l^2(l+k)^2R_{l}R_{l+k}} \left(\frac{R_{l+k}-R_{k}}{(l+k)^2-k^2}\right) + \frac{l_\mu k^\mu R_{k}}{l^2R_{l}(l+k)^2R_{l+k}}\left(\frac{R_{l+k}-R_{l}}{(l+k)^2-l^2}\right) - \frac{2 l_\mu k^\mu}{l^2R_{l}R_{l+k}}\nonumber\\
&& \times\left(\frac{R_{l+k}-R_{k}}{(l+k)^2-k^2}\right) \left(\frac{R_{l+k}-R_{l}}{(l+k)^2-l^2}\right) + \frac{2 k^2}{(l+k)^2R_{l}R_{l+k}} \left(\frac{R_{l}-R_{k}}{l^2-k^2}\right)^2 +\frac{k^2 l_\mu (l+k)^\mu}{l^2(l+k)^2R_{l}R_{l+k}}\nonumber\\
&& \times \left(\frac{R_{l}-R_{k}}{l^2-k^2}\right) \left(\frac{R_{l+k}-R_{k}}{(l+k)^2-k^2}\right) -\frac{2}{(l+k)^2-k^2}\left(\frac{R_{l+k}-R_{k}}{(l+k)^2-k^2} -\frac{R'_{k}}{\Lambda^2}\right)\frac{k^2}{l^2R_{l}} +\frac{2 l_\mu k^\mu}{l^2R_{l}}\nonumber\\
&& \times \left(\frac{R_{l}}{\left(l^2-(l+k)^2\right) \left(l^2-k^2\right)}+\frac{R_{l+k}}{\left((l+k)^2-l^2\right) \left((l+k)^2-k^2\right)} +\frac{R_{k}}{\left(k^2-l^2\right)\left(k^2-(l+k)^2\right)}\right)\nonumber\\
&& + \frac{1}{2\big((l+k)^2-l^2\big)}\left(\frac{2 R_{l+k} R'_{l+k} (l+k)^2}{\Lambda^2\big((l+k)^2 R_{l+k}^2+M_\varphi^2\big)} -\frac{2 R_l R'_l l^2}{\Lambda^2\big(l^2 R_l^2 + M_\varphi^2\big)} - \frac{1}{(l+k)^2 + M_\varphi^2} + \frac{1}{l^2+M_\varphi^2}\right.\nonumber\\
&&\left. +\frac{R_{l+k}^2}{(l+k)^2 R_{l+k}^2 + M_\varphi^2} - \frac{R_l^2}{l^2 R_l^2+M_\varphi^2} \right)\Biggr];\\
&& \vphantom{1}\nonumber\\
\label{General_G}
&& g(\xi_0,k/\Lambda) = \int \frac{d^4l}{(2\pi)^4} \Biggl[\frac{1}{2l^4}\Big(\frac{\xi_0}{K_l} - \frac{1}{R_l}\Big)\Big(R_{k+l} - \frac{2}{3} R_k\Big) - \frac{1}{2R_l l^2 (k+l)^4}\Big(\frac{\xi_0}{K_{k+l}}- \frac{1}{R_{k+l}}\Big)\nonumber\\
&& \times (k_\mu R_k + l_\mu R_l)^2 - \frac{R_k^2 k^2 l^\mu (k+l)_\mu}{4 l^4 (k+l)^4}\Big(\frac{\xi_0}{K_l} - \frac{1}{R_l}\Big) \Big(\frac{\xi_0}{K_{k+l}}-\frac{1}{R_{k+l}}\Big) \Biggr];\\
&& \vphantom{1}\nonumber\\
\label{General_H}
&& h(k/\Lambda) = \int \frac{d^4l}{(2\pi)^4} \frac{1}{\big((k+l)^2-l^2\big)}\Biggr[ - \frac{M^2 F'_{k+l}}{\Lambda^2 F_{k+l} \big((k+l)^2 F_{k+l}^2 + M^2\big)} + \frac{M^2 F'_{l}}{\Lambda^2 F_l \big(l^2 F_{l}^2 + M^2\big)}\nonumber\\
&& -\frac{1}{2(k+l)^2} + \frac{1}{2l^2} + \frac{F_{k+l}^2}{2\big((k+l)^2 F_{k+l}^2 + M^2\big)} - \frac{F_l^2}{2\big(l^2 F_l^2+M^2\big)} \Biggr],\qquad\quad
\end{eqnarray}

\noindent
where $R_k \equiv R(k^2/\Lambda^2)$, $K_k \equiv K(k^2/\Lambda^2)$, and $F_k \equiv F(k^2/\Lambda^2)$. Eqs. (\ref{General_Result}) --- (\ref{General_H}) are the main result of this paper. They give the expression for the one-loop polarization operator in the form of loop integrals and allow to simplify calculating sums of various multiloop superdiagrams which contain subdiagrams presented in Fig. \ref{Figure_Effective_Diagram}.

In the limit $k\to 0$ the derivatives of the functions $f$ and $h$ with respect to $\ln\Lambda$ can be presented as integrals of double total derivatives in the momentum space, so that

\begin{eqnarray}\label{Result_For_Vanishing_Momentum}
&& \frac{d\Pi}{d\ln\Lambda}\Big|_{k=0} = \pi \alpha_0 \int \frac{d^4l}{(2\pi)^4} \frac{d}{d\ln\Lambda} \Bigg( \frac{\partial}{\partial l^\mu} \frac{\partial}{\partial l_\mu}
\Big[\frac{C_2}{l^2} \ln\Big(1+\frac{M_\varphi^2}{l^2 R_l^2}\Big) + \frac{2 C_2}{l^2} \ln\Big(1+\frac{M_\varphi^2}{l^2}\Big)\qquad\nonumber\\
&& - \frac{T(R)}{l^2} \ln\Big(1+\frac{M^2}{l^2 F_l^2}\Big)\Big] + 8C_2 \Big(-\frac{1}{3 R_l l^4} + \frac{\xi_0}{3 K_l l^4} \Big) + O(\alpha_0,\lambda_0^2)\Bigg).
\end{eqnarray}

\noindent
This expression coincides with the corresponding result obtained in Ref. \cite{Aleshin:2016yvj}. This coincidence can be used for checking the correctness of the calculations. Calculating the integrals in Eq. (\ref{Result_For_Vanishing_Momentum}) we obtain

\begin{equation}
\frac{d\Pi}{d\ln\Lambda}\Big|_{k=0} = \frac{\alpha_0}{2\pi}\Big(3C_2-T(R)\Big) - \frac{\alpha_0 C_2(1-\xi_0)}{3\pi} + O(\alpha_0^2,\alpha_0\lambda_0^2).
\end{equation}

\noindent
Here the last term containing $(1-\xi_0)$ comes from the function $g$. Note that in Eq. (\ref{Result_For_Vanishing_Momentum}) the corresponding contribution is not given by an integral of total derivatives. Due to the presence of this contribution the function $\Pi$ appears to be gauge dependent. However, it is possible to remove all gauge dependent terms by choosing the Feynman gauge $\xi_0=1$ and imposing the condition $R(x) = K(x)$, because in this case $g = 0$.

It is interesting to find the explicit expressions for the functions $f(k/\Lambda)$ and $h(k/\Lambda)$ in the limit $k/\Lambda \to 0$. In the Appendix we demonstrate that the former function can be written as

\begin{equation}\label{Explicit_F}
f(k/\Lambda) = -\frac{3}{16\pi^2} \Big(\ln \frac{M_\varphi}{k} + 1 + o(1)\Big) = -\frac{3}{16\pi^2} \Big(\ln \frac{\Lambda}{k} + \ln a_\varphi + 1 + o(1)\Big),
\end{equation}

\noindent
where $o(1)$ denotes terms vanishing in the limit $k/\Lambda\to 0$.

Also it is possible to find the asymptotic behaviour of the function $h(k/\Lambda)$. Again, this is done in the Appendix. The result has the form

\begin{equation}\label{Explicit_H}
h(k/\Lambda) = \frac{1}{16\pi^2} \Big(\ln \frac{M}{k} + 1 + o(1)\Big) = \frac{1}{16\pi^2} \Big(\ln \frac{\Lambda}{k} +\ln a + 1 + o(1)\Big).
\end{equation}

\noindent
Note that Eqs. (\ref{Explicit_F}) and (\ref{Explicit_H}) include not only the logarithmically divergent terms, but also the finite constants. Certainly, the logarithmically divergent terms are removed by performing the one-loop renormalization (see, e.g., \cite{Bogolyubov:1980nc}), which we will not discuss here.

\section{Conclusion}
\hspace*{\parindent}

In this paper we present the expression for the one-loop polarization operator of the quantum gauge superfield in the one-loop approximation in the case of using the BRST-invariant version of the higher covariant derivative regularization. The main result of this paper given by Eq. (\ref{General_Result}) is divided into three parts: the part corresponding to the supersymmetric Yang--Mills theory without matter in the minimal gauge ($\xi_0=1$ and $K(x)=R(x)$) is included into the function $f$, the function $g$ contains all gauge dependent terms, and the matter contributions are collected in the function $h$. All these functions are given by complicated integrals which can be written as sums of logarithmically divergent terms, finite constants, and terms vanishing in the limit $k\to 0$. In the minimal gauge the logarithmical divergences are determined by the integrals of double total derivatives, in agreement with Ref. \cite{Aleshin:2016yvj}. In this paper we also calculate constant terms in the minimal gauge for an arbitrary choice of the higher derivative regulators $R(x)$ and $F(x)$. These finite terms appear to be independent of $R(x)$ and $F(x)$, because the contributions depending on these functions are given by integrals of total derivatives in the momentum space.

The expression for the polarization operator (\ref{General_Result}) may considerably simplify calculations of various Green functions in higher orders of the perturbation theory, because it allows summing multiloop diagrams containing the subdiagram presented in Fig. \ref{Figure_Effective_Diagram}. We hope to use it for doing various two- and three-loop calculations in non-Abelian supersymmetric theories regularized by higher derivatives and investigating the possibility to present the $\beta$-function in the NSVZ form in higher orders of the perturbation theory.

\section*{Acknowledgements}
\hspace*{\parindent}

The work of A.K. and M.S. was supported by the Foundation "BASIS", grant No. 17-11-120.

The authors are very grateful to A.L.Kataev for valuable discussions.

\section*{Appendix: Asymptotic behaviour of the functions $f(k/\Lambda)$ and $h(k/\Lambda)$}
\hspace{\parindent}

Let us find the asymptotic behaviour of the function $f(k/\Lambda)$ in the limit $k/\Lambda \to 0$. For this purpose we present the expression (\ref{General_F}) in the form

\begin{equation}
f(k/\Lambda) = f_1(k/\Lambda) + f_2(k/\Lambda),
\end{equation}

\noindent
where

\begin{eqnarray}\label{Expression_For_F1}
&&\hspace*{-9mm} f_1(k/\Lambda) \equiv - \frac{3}{2} \int \frac{d^4l}{(2\pi)^4} \left(\frac{1}{l^2(l+k)^2} -\frac{1}{(l^2+M_{\varphi}^2)((l+k)^2+M_{\varphi}^2)}\right) \nonumber\\
&&\hspace*{-9mm} = -\frac{3}{16\pi^2}\Bigg(\ln\frac{M_\varphi}{k} + \sqrt{1+\frac{4M_\varphi^2}{k^2}}\, \mbox{arctanh} \sqrt{\frac{k^2}{k^2+4 M_\varphi^2}}\, \Bigg) =  -\frac{3}{16\pi^2} \Big(\ln \frac{M_\varphi}{k} + 1 + o(1)\Big)
\end{eqnarray}

\noindent
and $f_2(k/\Lambda)$ includes all remaining terms of Eq. (\ref{General_F}). (The derivation of Eq. (\ref{Expression_For_F1}) can be found, e.g., in \cite{Soloshenko:2002np,Soloshenko:2003sx}.) Using Eq. (\ref{General_F}), in the limit $k/\Lambda\to 0$ we obtain

\begin{equation}\label{F_AT_0}
f_2(0) = \int \frac{d^4l}{(2\pi)^4} \frac{d}{dl^2} \left(\frac{1}{2(l^2 + M_\varphi^2)} - \frac{R_l^2}{2(l^2 R_l^2 + M_\varphi^2)} + \frac{M_\varphi^2 R_l'}{\Lambda^2 R_l\big(l^2 R_l^2 + M_\varphi^2\big)}\right).
\end{equation}

\noindent
Evidently, the integrand does not depend on the angular variables. Therefore, after integrating over the angles, $d^4l \to \pi^2 l^2 dl^2$. Introducing a new variable $x\equiv l^2/\Lambda^2$ and integrating by parts, after some transformations the considered expression can be written as an integral of a total derivative,

\begin{eqnarray}
&& f_2(0) = -\frac{1}{16\pi^2} \int\limits_{0}^\infty dx\left(\frac{a_\varphi^2 R'(x)}{R(x)\big(x R^2(x) + a_\varphi^2\big)} +\frac{1}{2(x+a_\varphi^2)} - \frac{R^2(x)}{2\big(x R^2(x) + a_\varphi^2\big)}\right)\qquad\nonumber\\
&& = \frac{1}{32\pi^2} \int\limits_{0}^\infty dx\, \frac{d}{dx} \ln \frac{1+a_\varphi^2/\big(x R^2(x)\big)}{1+a_\varphi^2/x} = 0.
\end{eqnarray}

\noindent
Therefore, in the limit $k/\Lambda\to 0$

\begin{equation}
f(k) = -\frac{3}{16\pi^2} \Big(\ln \frac{M_\varphi}{k} + 1 + o(1)\Big).
\end{equation}

The asymptotic behaviour of the function $h(k/\Lambda)$ can be found by a similar method. First, we present this function in the form

\begin{eqnarray}
h(k/\Lambda) = h_1(k/\Lambda) + h_2(k/\Lambda),
\end{eqnarray}

\noindent
where

\begin{eqnarray}
&&\hspace*{-10mm} h_1(k/\Lambda) \equiv \frac{1}{2} \int \frac{d^4l}{(2\pi)^4} \left(\frac{1}{l^2(l+k)^2} -\frac{1}{(l^2+M^2)((l+k)^2+M^2)}\right)\nonumber\\
&&\quad\qquad\qquad\qquad\qquad\qquad\qquad\qquad\qquad\qquad\qquad\qquad =  \frac{1}{16\pi^2} \Big(\ln \frac{M}{k} + 1 + o(1)\Big);\\
&&\hspace*{-10mm} h_2(k/\Lambda) = \int \frac{d^4l}{(2\pi)^4} \frac{1}{\big((k+l)^2-l^2\big)}\left(\vphantom{\frac{1}{2}}\right. - \frac{M^2 F'_{k+l}}{\Lambda^2 F_{k+l} \big((k+l)^2 F_{k+l}^2 + M^2\big)} + \frac{M^2 F'_{l}}{\Lambda^2 F_l \big(l^2 F_{l}^2 + M^2\big)} \nonumber\\
&&\hspace*{-10mm} -\frac{1}{2\big((k+l)^2+M^2\big)} + \frac{1}{2\big(l^2+M^2\big)} + \frac{F_{k+l}^2}{2\big((k+l)^2 F_{k+l}^2 + M^2\big)} - \frac{F_l^2}{2\big(l^2 F_l^2+M^2\big)} \left.\vphantom{\frac{1}{2}}\right).
\end{eqnarray}

\noindent
Unlike $h(k/\Lambda)$, the function $h_2(k/\Lambda)$ has the limit at $k/\Lambda\to 0$,

\begin{equation}
h_2(0) = \int \frac{d^4l}{(2\pi)^4} \frac{d}{dl^2} \left(- \frac{M^2 F'_{l}}{\Lambda^2 F_l \big(l^2 F_{l}^2 + M^2\big)} - \frac{1}{2\big(l^2+M^2\big)} + \frac{F_l^2}{2\big(l^2 F_l^2+M^2\big)} \right).
\end{equation}

\noindent
Comparing this expression with Eq. (\ref{F_AT_0}) and making the transformations similar to the ones described above, we obtain

\begin{equation}
h_2(0) = -\frac{1}{32\pi^2} \int\limits_{0}^\infty dx\, \frac{d}{dx} \ln \frac{1+a^2/\big(x F^2(x)\big)}{1+a^2/x} = 0,
\end{equation}

\noindent
where $x\equiv l^2/\Lambda^2$, as earlier. Note that the expression $h_2(0)$ (similarly to the expression $f_2(0)$) is given by a vanishing integral of a total derivative.

\end{document}